\documentclass[abbrv,aps,prb,twocolumn,showpacs,preprintnumbers,amsmath,amssymb,nofootinbib,superscriptaddress]{revtex4-1}

\usepackage{graphicx}
\usepackage{color}
\usepackage[ansinew]{inputenc}

\newcommand{\ang}{\ensuremath{\rm{\AA}}}

\newcommand{\bra}[1]{\ensuremath{\langle #1|}}
\newcommand{\ket}[1]{\ensuremath{|#1\rangle }}

\begin{document}

\title{Transition metal ad-atoms on graphene: Influence of local Coulomb interactions on chemical bonding and magnetic moments}
\author{T. O. Wehling}
\email{twehling@physnet.uni-hamburg.de}
\author{A. I. Lichtenstein}
\affiliation{1.\ Institut für Theoretische Physik, Universität Hamburg, 
D-20355 Hamburg, Germany}
\author{M. I. Katsnelson}
\affiliation{Institute for Molecules and Materials, Radboud University 
Nijmegen, NL-6525 AJ Nijmegen, The Netherlands}

\begin{abstract}
We address the interaction of graphene with $3d$ transition metal adatoms and the formation of localized magnetic moments by means of first-principles calculations. By comparing calculations within the generalized gradient approximation (GGA) to GGA+U we find that the electronic configuration and the adsorption geometries can be very sensitive to effects of local Coulomb interactions $U$ in the transition metal $d$-orbitals. We find high-spin configurations being favorable for Cr and Mn adatoms independent of the functional. For Fe, Co and Ni different electronic configurations are realized depending on the value of the local Coulomb interaction strength $U$. Chemical control over the spin of the adatoms by hydrogenation is demonstrated: NiH and CoH adsorbed to graphene exhibit spin $S=1/2$ and $S=1$, respectively.
\end{abstract}

\maketitle

\section{Introduction}
Graphene --- a single layer of C-atoms arranged to a honeycomb lattice --- has developed from a theoretical model system \cite{Wallace-1947,McClure_56} to experimental reality \cite{Novoselov_science2004}. It is stabilized by a network of $sp^2$ $\sigma$-bonds and conjugated $\pi$ bonds, which make this material chemically inert. Its low energy electronic structure is determined by two $\pi$ bands intersecting linearly in the corners of the Brillouin zone. The Fermi level of the pristine material lies at these crossing points --- called Dirac points. As a consequence, graphene exhibits Dirac fermion excitations and a linearly vanishing density of states (DOS) at low energies. Carrier concentrations have been proven tunable --- both for electrons and holes --- by more than an order of magnitude by chemical \cite{SchedinGassensors,NanoLettAds,AdamFuhrerIshigamiCargedImp08} as well as electrostatic doping \cite{Novoselov_science2004}. It can be anticipated that this control over the electronic density (of states) implies a general tunability of electron correlation phenomena in graphene.

Adsorbates carrying a magnetic moment, like $3d$ transition metal adatoms, and binding to graphene present a chemical route to localized magnetic moments in graphene and possibly allow us probing tunable many body physics in this material. 
Various studies have addressed the adsorption of $3d$ adatoms by means of density functional (DFT) calculations \cite{Kawazoe04,Cohen_PRB08,Mao_Co_graphene_09,Tok_PRB09,Longo_PRB10,Co_Fano_PRB10,Cao_PRB10,CoGraphene_Kondo_PRB10,Yazyev_PRB10,Jacob_Co_Graphene10,Co_adat_CohenPRB11,Ho_PRB11,Nakada11}. 

Many $3d$-adatoms including Fe, Co and Ni were predicted to adsorb in the center of a graphene hexagon (h-site) \cite{Cohen_PRB08,Mao_Co_graphene_09}. At h-sites, the coordination number is maximized for the adatom. Its $d$-orbitals with non-zero $z$-component of the orbital angular momentum ($l_z\neq0$) couple to the low energy graphene states, whereas the $s$-orbital of the adatoms remains decoupled from the low energy states a this site \cite{Co_Fano_PRB10,CoGraphene_Kondo_PRB10}. Away from the h-site position, the $s$-orbital can hybridize with low energy graphene states but forming a $\sigma$-bond with one of the carbon atoms requires energy for breaking graphene's $sp^2$-network \cite{MonoImp09}. Moreover, the reduced coordination decreases the $d$-electron contribution to bonding. The adsorption site of adatoms relies hence on balancing the contributions to chemical bonding from the $s$- against those from the d-orbitals. The spatial extent of the latter orbitals is strongly influenced by the local Coulomb interaction. Therefore, the energy gain upon bonding can be very sensitive to effects of on-site Coulomb repulsion. 

In this paper, we investigate how the interaction of $3d$ transition metal adatoms with graphene is affected by local Coulomb interactions and consider Cr, Mn, Fe, Co, and Ni atoms adsorbed to graphene. In section \ref{sec:DFT_details} details of our DFT calculations are given. In section  \ref{sec:3d_ad_atoms}, we present our results on adsorption geometries and the electronic structure of $3d$ adatoms on graphene. The influence of local Coulomb interactions is discussed. In section \ref{sec:hydro_adat}, we show that the magnetic state of adatoms on graphene is very sensitive to their chemical environment by considering hydrogenated Co and Ni adatoms on graphene. Conclusions are given in section \ref{sec:conclusions}.

\section{Computational method}
\label{sec:DFT_details}
For an ab-initio description of the transition metal adatoms on graphene we performed DFT calculations on $3\times 3$ and $4\times 4$ graphene supercells containing one adatom using the Vienna Ab Initio Simulation Package
(VASP) \cite{Kresse:PP_VASP} with the projector augmented wave (PAW) \cite{Bloechl:PAW1994,Kresse:PAW_VASP}
basis sets. To judge the role of local Coulomb interactions in the $3d$ orbitals of the adatoms, we employed a generalized gradient
approximation functional (GGA) \cite{Perdew:PW91} as well as GGA+U with $U=4$\,eV, $J=0.9$\,eV, which is typical value for 3d-transition metal impurities in metallic hosts \cite{Anisimov_PRB94,Anisimov_Lichtenstein_LDAU_97}. In addition, LDA calculations were performed. The calculations were done with plane wave cut-offs $800$\,eV and $928$\,eV for the GGA and LDA functionals, respectively. We obtained fully relaxed structures for all of these functionals with the forces acting on each atom being less than $0.02$\,eV$\ang^{-1}$.

The binding energies $E_b$ reported in the following were obtained using the standard formula $E_b=E_{\rm g}+E_{\rm ads}-E_{\rm g-ads}$, where $E_{\rm g-ads}$, $E_{\rm g}$ and $E_{\rm ads}$ are, respectively, the total energies of the graphene-adsorbate system, a pristine graphene sheet and the isolated adsorbate. As isolated atoms with open electronic shells are strongly correlated systems, the GGA or GGA+U values for $E_{\rm ads}$ can be unreliable and it is hence most meaningful to consider binding energy differences where $E_{\rm ads}$ cancels out.

\section{Magnetism and structure of $3d$ adatoms on graphene}
\label{sec:3d_ad_atoms}
\begin{figure*}
\centering
 \includegraphics[width=.98\linewidth]{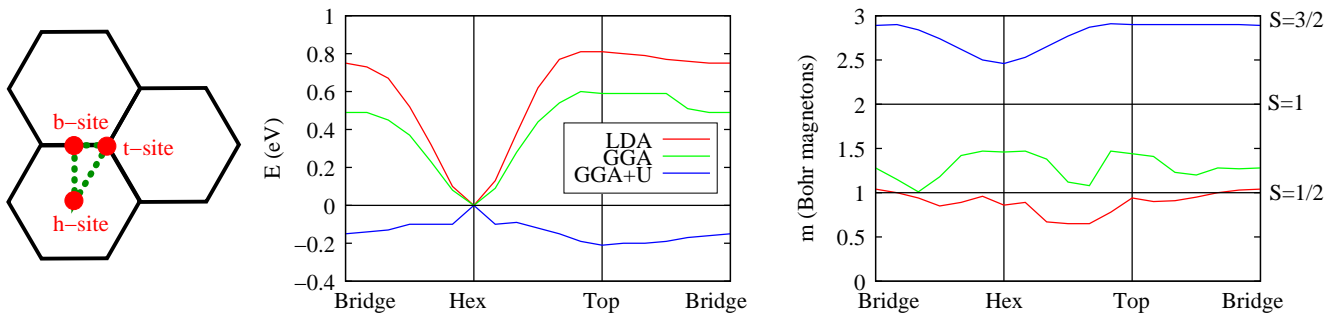}
 \caption{\label{fig:Co_graphene_energy_magnet} (Color online) Left: Schematic illustration high symmetry adsorption sites for Co adatoms on graphene (red) and paths connecting these sites (green). Middle: Total energy, $E$, of graphene with a Co adatom as function of the Co position along the path depicted in the left panel. Energies obtained with GGA, LDA, and GGA+U ($U=4$\,eV, $J=0.9$\,eV) are shown. The energy for Co at an h-site is defined as $E=0$. Right: Total magnetic moment of the graphene supercell with adsorbed Co as function of the Co position.}
\end{figure*}
We recapitulate the example of Co which has been subject of prior studies \cite{Mao_Co_graphene_09,Co_Fano_PRB10,CoGraphene_Kondo_PRB10,Jacob_Co_Graphene10,Co_adat_CohenPRB11} and then generalize to Cr, Mn, Fe and Ni adatoms on graphene.

To judge which adsorption geometries are possible, we firstly relaxed Co adatoms adsorbed to high-symmetry sites the graphene lattice: Co above the center of a hexagon (h-site), Co on top of a C-atom (t-site) and above the middle of the bridge (b-site). Using a $4\times 4$ graphene supercell, $100$\,meV first order Methfessel-Paxton smearing and a $6\times 6$ k-mesh for sampling the supercell Brillouin zone, we calculated to total-energy and the total magnetic moment of the supercell for the Co atom being moved along the path connecting the relaxed adsorption geometries\footnote{The paths were obtained by linearly interpolating between the relaxed adsorption geometries and using 5 images in between each of the relaxed geometries.}.
As can be seen from Fig. \ref{fig:Co_graphene_energy_magnet}, LDA as well as GGA predict Co adsorbing to a h-site with t-site or b-site adsorption being $\gtrsim 0.5$\,eV higher in energy. In all cases, the supercell magnetic moment is near $m\approx 1\,\mu_B$ suggesting an electronic configuration of Co close to spin $S=1/2$. These results are in agreement with prior studies \cite{Mao_Co_graphene_09,Co_Fano_PRB10,CoGraphene_Kondo_PRB10}.


Orbital dependent on-site Coulomb repulsion, as accounted for in GGA+U in a mean field manner, affects the predicted configuration of Co on graphene drastically: With $U=4$\,eV, we find that the potential energy landscape is reversed as compared to the GGA / LDA prediction (see Fig. \ref{fig:Co_graphene_energy_magnet} middle). For $U=4$\,eV, $J=0.9$\,eV the global minimum energy is found for Co in a  high spin state with $S=3/2$ on a t-site, which is $0.2$\,eV and $0.08$\,eV lower in energy than the h- and b-sites, respectively. 

\begin{figure*}
\centering
\includegraphics[width=0.8\linewidth]{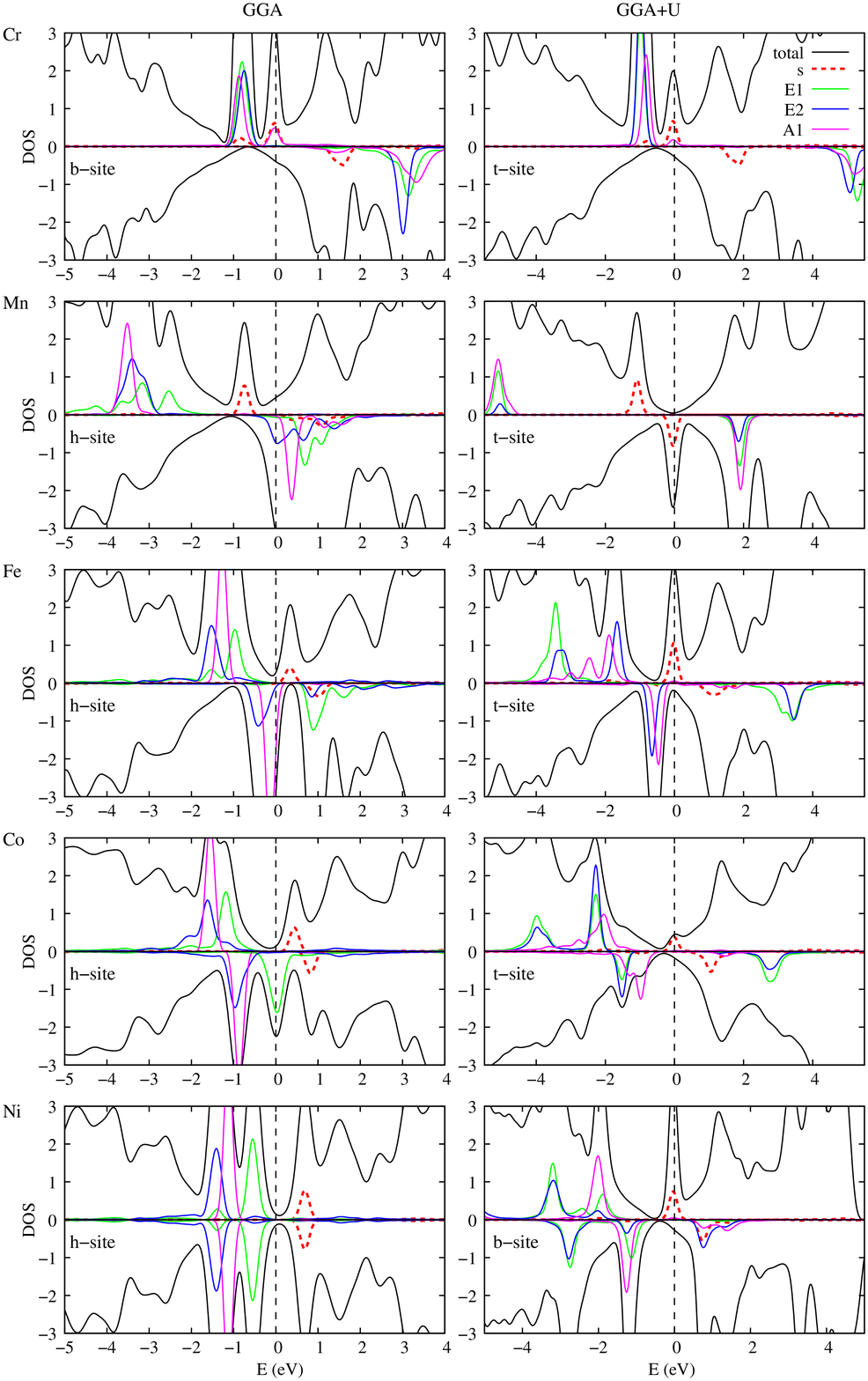}
\caption{(Color online) Local orbitally and spin resolved density of states at the adatoms (Cr, Mn, Fe, Co, and Ni from top to bottom) and total density of states of the supercells obtain with GGA (left column) as well as GGA+U (right column, $U=4$\,eV, $J=0.9$\,eV). Spin-up is plotted on the positive ordinate, spin-down on the negative ordinate. The GGA+U high spin solution show for Ni is metastable at $U=4$\,eV. All other LDOS shown here belong to structures with lowest total energy.}
\label{fig:DOS_GGA_GGA+U_adatoms}
\end{figure*}

The local density of states depicted in Fig. \ref{fig:DOS_GGA_GGA+U_adatoms} show that the electronic configuration of Co markedly changes upon increasing the local Coulomb interaction $U$. In the GGA low spin solution the Co $4s$-orbital is unoccupied and approximately one hole resides in the Co d-orbitals with E1 symmetry ($d_{xz}$, $d_{yz}$), i.e. Co is close to a $3d^9 4s^0$ configuration. In contrast, at $U=4$\,eV Co appears in a $3d^8 4s^1$ configuration. 

This change in the electronic properties comes along with a change in equilibrium heights of the Co adatoms: With the increase of the Co spin from $\sim 1/2$ to $\sim 3/2$, the equilibrium height of Co above the sheet increases from $1.5\,\ang$ to $1.8\,\ang$ for the adatoms at an h-site as well as from $1.8\,\ang$ to $2.2\,\ang$ for the adatoms placed at the t-site. This observation is in agreement with Ref. \onlinecite{Jacob_Co_Graphene10}, where the B3LYP hybrid functional has been employed. The change in geometry can be traced back to the more extended Co $4s$-orbital becoming occupied, when increasing the Hubbard-U and the Co spin. Balancing electrons between the Co $3d$ and $4s$-orbitals is hence decisive for determining the electronic properties and the adsorption geometry of Co on graphene.

This balancing of electrons between the $3d$ and $4s$-orbitals is related with the interconfiguration energies like $\Delta E_{\rm ic}=E(4s^1 3d^{n-1})-E(4s^2 3d^{n-2})$. For free $3d$-atoms it is known that semilocal functionals like LDA or GGA tend to underestimate $\Delta E_{\rm ic}$ and generally favor $4s$ occupancies lower than experimentally observed \cite{Kresse_PRB97}. For Co adatoms on graphene, it is not yet clear whether the LDA/GGA configuration with Co close to a $3d^9 4s^0$ or the GGA+U $3d^8 4s^1$ configuration is closer to the experiment or whether even both configurations might be realized. The example of Co adatoms, however, demonstrates that for a general understanding of $3d$ transition metal adatoms on graphene comparing semilocal LDA/GGA type functionals to GGA+U can be useful. 

To this end, we performed GGA as well as GGA+U ($U=4$\,eV, $J=0.9$\,eV) calculations for Cr, Mn, Fe, Co, and Ni on graphene (one adatom per $3\times 3$ graphene supercell), which allowed us to obtain fully relaxed geometries, total-energies, and magnetic properties. The results are summarized in Table \ref{tab:3d_geom_mag}.
\begin{table*}
\centering
\begin{minipage}{0.33\linewidth}
\begin{tabular}{|l|c|c|c||c|c|c|}
\hline
\multicolumn{7}{|c|}{Magnetic moment ($\mu_B$)}\\
\hline
&\multicolumn{3}{c||}{GGA}&\multicolumn{3}{c|}{GGA+U}\\ 
\hline
&	t &	b	&h&	t& 	b&	h\\
\hline
Cr&	5.9&	5.8&	5.6&	5.9&	5.9&	5.9\\
Mn&	5.1&	5.1&	5.2&	5.0&	5.0&	5.0\\
Fe&	4.1&	4.1&{\color{red}	2.0}&	3.9&	3.9&	3.6\\
Co&	{\color{red}1.1}&	{\color{red}1.1}&	{\color{red}1.1}&	2.9&	2.9&	2.6\\
Ni&	{\color{red}0.0}&	{\color{red}0.0}&	{\color{red}0.0}&	{\color{red}0.0}&	{\color{red}0.0}&	{\color{red}0.0}\\
\hline
\end{tabular}
\end{minipage}\begin{minipage}{0.33\linewidth}
\begin{tabular}{|l|c|c|c||c|c|c|}
\hline
\multicolumn{7}{|c|}{Height above sheet ($\ang$)}\\
\hline
&\multicolumn{3}{c||}{GGA}&\multicolumn{3}{c|}{GGA+U}\\ 
\hline
&	t &	b	&h&	t& 	b&	h\\
\hline
Cr&	2.4&	2.3&	2.1&	2.6&	2.6&	2.9\\
Mn&	2.2&	2.2&	2.1&	3.7&	3.8&	3.8\\
Fe&	2.2&	2.2&	{\color{red}1.5}&	2.2&	2.3&	2.0\\
Co&	{\color{red}1.9}&	{\color{red}1.8}&	{\color{red}1.5}&	2.2&	2.2&	1.8\\
Ni&	{\color{red}1.9}&	{\color{red}1.8}&	{\color{red}1.6}&	{\color{red}1.8}&	{\color{red}1.8}&	{\color{red}1.5}\\
\hline
\end{tabular}
\end{minipage}
\begin{minipage}{0.33\linewidth}
\begin{tabular}{|l|c|c|c||c|c|c|}
\hline
\multicolumn{7}{|c|}{Binding energy (eV)}\\
\hline
&\multicolumn{3}{c||}{GGA}&\multicolumn{3}{c|}{GGA+U}\\ 
\hline
&	t &	b	&h&	t& 	b&	h\\
\hline
Cr&	0.49&	0.50&	0.47&	0.34&	0.34&	0.33\\
Mn&	0.35&	0.34&	0.42&	0.30&	0.28&	0.28\\
Fe&	0.24&	0.22&	{\color{red}0.71}&	0.29&	0.25&	0.27\\
Co&	{\color{red}0.75}&	{\color{red}0.88}&	{\color{red}1.35}&	0.62&	0.56&	0.41\\
Ni&	{\color{red}1.44}&	{\color{red}1.52}&	{\color{red}1.72}&	{\color{red}0.74}&	{\color{red}0.82}&	{\color{red}0.97}\\
\hline
\end{tabular}
\end{minipage}
\caption{Magnetic moment of supercell, height above graphene sheet and energy relative h-site for $3d$ transition metal adatoms as obtained from GGA and GGA+U with $U=4$\,eV and $J=0.9$\,eV. High-spin solutions are colored back, low-spin solution red (grey).}
\label{tab:3d_geom_mag}
\end{table*}
Two classes of adsorbed electronic configurations are obtained: Those with the supercell magnetic moment close to the free atom moment (high-spin configuration, colored black) and those exhibiting a by $\sim 2\mu_B$ smaller magnetic moment than the free atoms (low-spin configuration colored red (grey)). The latter solutions become increasingly favorable for the later $3d$ elements: In agreement with Refs.  \onlinecite{Cohen_PRB08,Mao_Co_graphene_09,Co_Fano_PRB10,CoGraphene_Kondo_PRB10,Longo_PRB10,Cao_PRB10} our GGA calculations yield Fe in low-spin configuration at an h-site, as well as Co and Ni in low-spin configuration at all sites. 

In all cases with stable low-spin configuration, h-site adsorption is favored. The LDOS (Fig. \ref{fig:DOS_GGA_GGA+U_adatoms}) shows that in all low-spin configurations $4s$-orbitals of the adatoms are unoccupied with electrons being transferred to the $3d$-orbitals. In the low spin h-site configurations obtained for Fe, Co, and Ni the $d_{xz}$- and $d_{yz}$-orbitals ($E_1$ type orbitals) are the $3d$-orbitals with the highest energy. For Fe (Co), GGA predicts a spin of $S=1$ ($S=1/2$) due to two (one) holes in the $E_1$ type orbitals, which are potentially most important for spin fluctuations in both cases.


As the example of isolated Co adatoms demonstrated, local orbital-dependent Coulomb interaction can strongly affect the balance of $3d$ and $4s$ electrons and consequently change the adsorption geometry as well as the magnetic state of the adatom. As in the case of Co, the low-spin solution of Fe at an h-site becomes unstable for $U=4$\,eV. At $U=4$\,eV there exist only high-spin solutions for Cr, Mn, Fe, and Co. 

Only Ni remains in the low-spin configuration also for $U=4$\,eV. There are however metastable GGA+U high-spin solutions also for Ni with magnetic moments $m=1.9\mu_B$ and $m=1.6\mu_B$ for Ni at b- and h-sites, respectively. These high-spin solutions are, respectively, $0.27$\,eV and $0.35$\,eV higher in energy than the low-spin h-site configuration. At $U=6$\,eV, the b-site high spin configuration becomes energetically most favorable.

The LDOS at the adatoms and the total DOS depicted in Fig. \ref{fig:DOS_GGA_GGA+U_adatoms} illustrate bonding mechanisms depending on the electronic configurations of the adsorbates. For all species in low-spin configuration there is strong hybridization between the $d$-orbitals and the graphene bands leading to the supercell Dirac point to being no more recognizable as discussed for Fe in Ref. \onlinecite{Cohen_PRB08}. 

For Cr, both, GGA and GGA+U predict high-spin states with similar LDOS near the Fermi level ($E=0$) and the Dirac point being visible for the minority spin spin states at $\approx -0.65$\,eV and $-0.52$\,eV, respectively. Given the $3\times 3$ supercell size this indicates ionic bonding with Cr donating on the order of $0.1-0.2$ electrons to the graphene bands.\footnote{The exact amount of charge transfer can depend on the adatom concentration / supercell size. The numbers given, here, should be considered qualitatively.} Similarly, GGA predicts ionic bonding of Mn on graphene with the Dirac point shift indicating $0.3-0.5$ electrons being transferred to the graphene bands. GGA+U yields Mn close to the atomic $3d^5 4s^2$ electronic configuration and much weaker bonding to graphene. For Fe, various resonances in the vicinity of the Fermi level hinder to distinguish between graphene bands and adatom induced states making an estimate of doping due to Fe ambiguous. The supercell magnetic moment $m=3.9\,\mu_B$ in the high-spin state suggests Fe acting as donor with $\approx 0.1$e being transfered to graphene.

The LDOS of the $4s$-orbitals of the adatoms consist mainly of a narrow peak and indicate at $4s$-electrons remaining localized upon adsorption of the atoms on graphene. In many high-spin solutions (Cr (GGA and GGA+U), Mn (GGA), Fe (GGA+U), Co (GGA+U), and Ni (GGA+U)), the $4s$ resonance of the adatoms is occupied by (almost) one electron and close to the Fermi level. In these cases, the impurity $4s$-orbitals contribute a magnetic moment of $1\mu_B$. Hence, when discussing the low energy physics (e.g. the Kondo effect) of adatoms in high-spin configurations on graphene, the $4s$-orbital has to be taken into account, as it is potentially important for charge or spin-fluctuations of the magnetic adatom.


The coexistence of high- and low-spin solutions for Fe, Co, and Ni suggests that the spin of transition metal adatoms can be tunable. In the following section, we demonstrate that hydrogenation presents chemical means to control the magnetic properties of adatoms on graphene with the examples of Co and Ni. 

\section{Hydrogenated Co and Ni adatoms}
\label{sec:hydro_adat}
\begin{figure}
\begin{minipage}{0.49\linewidth}
\includegraphics[width=0.98\linewidth]{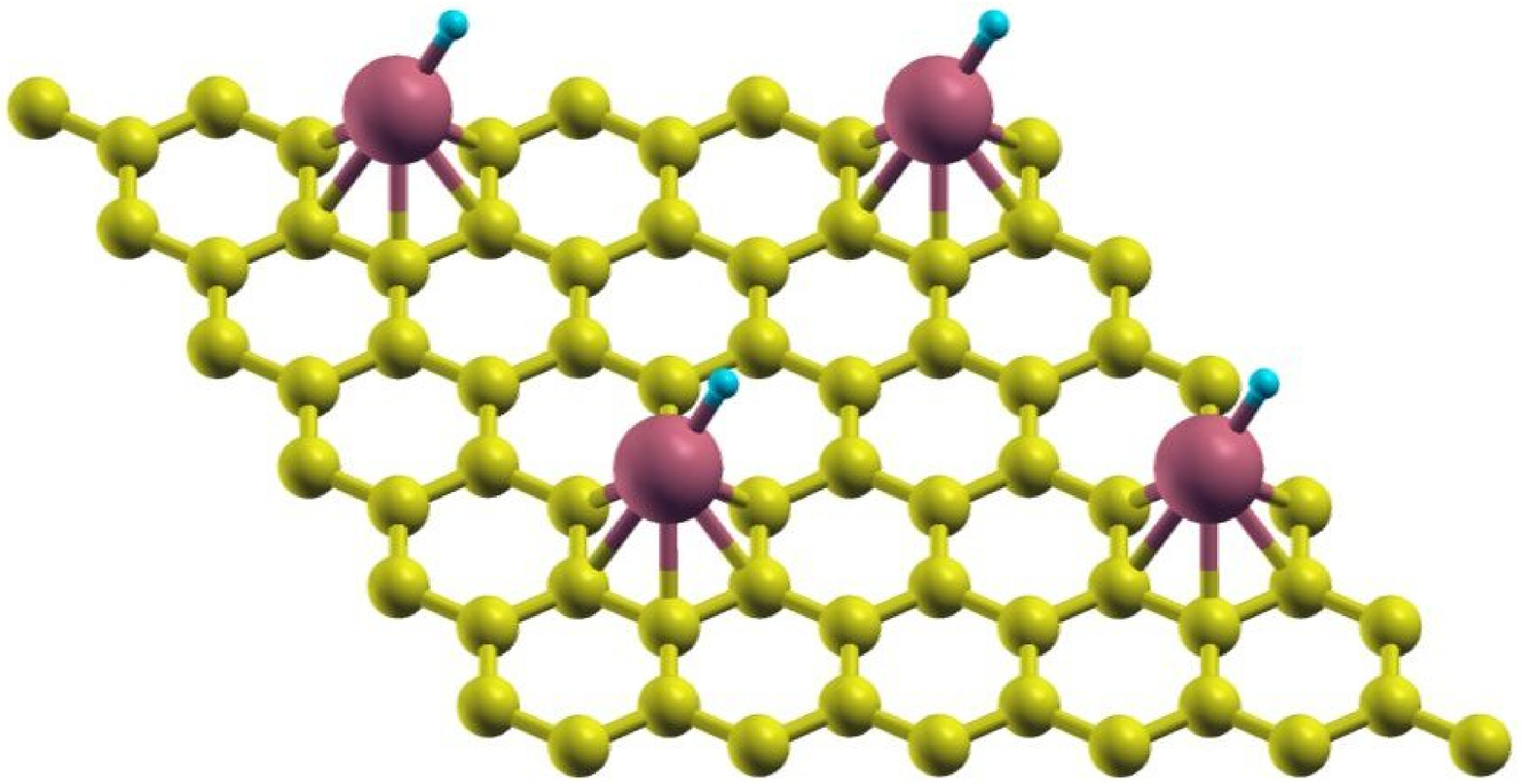}%
\end{minipage}
\begin{minipage}{0.49\linewidth}
\includegraphics[width=0.98\linewidth]{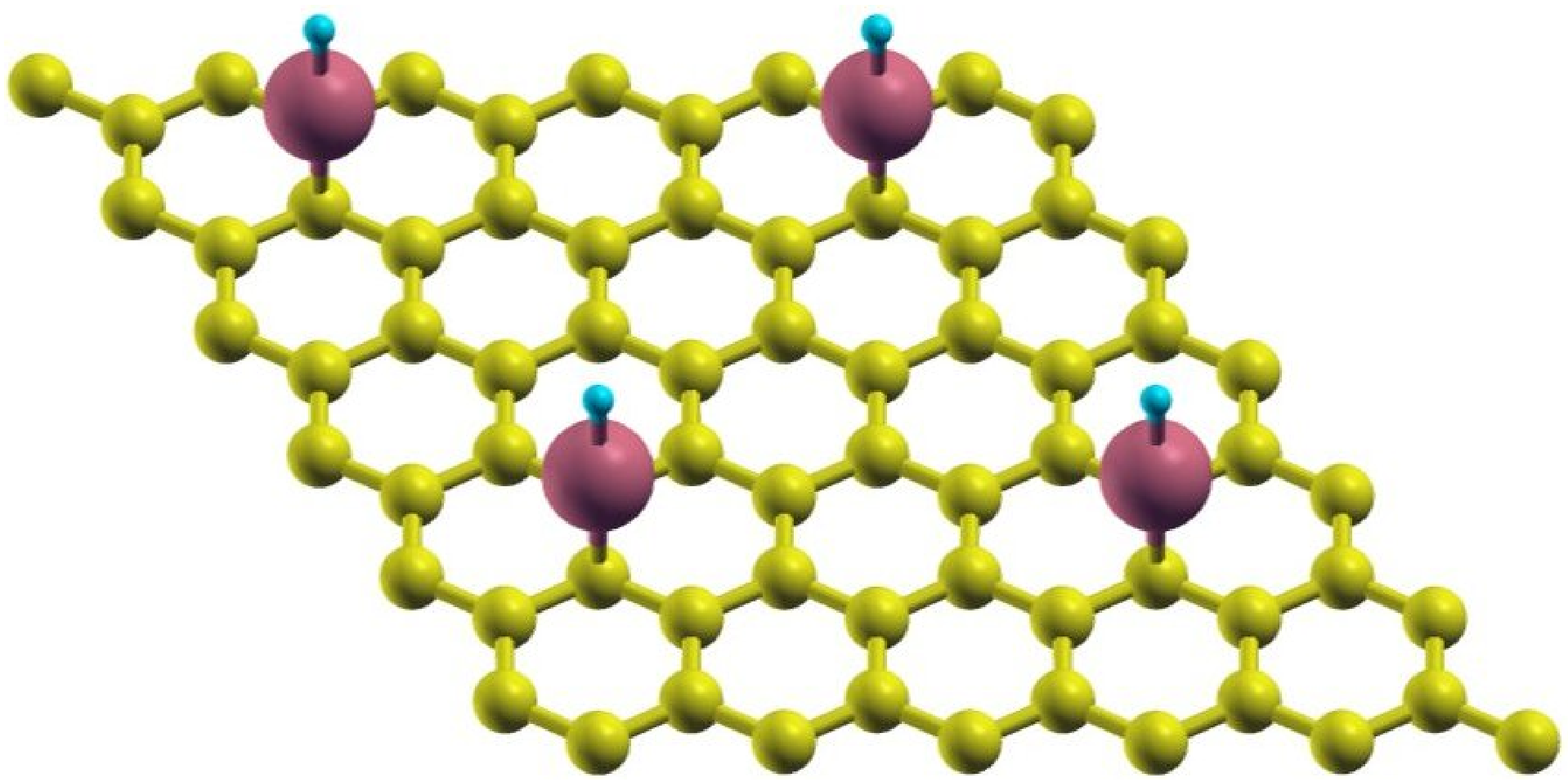}%
\end{minipage}
\caption{(Color online) Relaxed adsorption geometries of CoH on graphene as predicted by GGA (left) and GGA+U (right). For NiH, both, GGA and GGA+U yield h-site adsorption with virtually the same structure as obtained for CoH in GGA (left).}%
\label{fig:CoH_NiH_struct}%
\end{figure}

Transition metal $s$-orbitals are, in general, highly susceptible to chemical bonding. In bulk $3d$-elements, the $4s$-orbital is completely delocalized causing metallic bonding and gives rise to a conduction electron bandwidth on the order of some $10$\,eV. Even in the case of reduced coordination, e.g. for Co adatoms on metal surfaces like Cu (111) (see e.g. \onlinecite{CoCu_n_PRL08}), the characteristic energy band width of the $4s$-orbital is on the order of $10$\,eV. Similarly, for transition metal adatoms on insulating surfaces like CuN,\cite{Rudenko_09} covalent bonds cause distribution of $4s$ spectral weight over an energy range on the order of $10$\,eV.

For adatoms on graphene, the situation turned out to be very different (see section \ref{sec:3d_ad_atoms}), where the $4s$-orbital forms no strong covalent bond with a carbon atom underneath. This ``frustration'' of the transition metal $4s$-orbital can be resolved by chemical bonds to other species. Hydrogen is available in many experimental situations and consequently particularly important. Here, we consider the interaction of H with Co and Ni adatoms, where the $4s$-orbital is predicted by GGA to be unoccupied. 

Our structural relaxations show that H forms a chemical bond with the Ni and Co adatoms leading to a CoH bond length of $1.54\,\ang$ ($1.55\,\ang$) and a NiH bond length of $1.49\,\ang$ ($1.51\,\ang$) in GGA (GGA+U; $U$ acting on transition metal d-orbitals; $U=4$\,eV, $J=0.9$\,eV). The fully relaxed adsorption geometries are shown in Fig. \ref{fig:CoH_NiH_struct}.
For NiH both, GGA and GGA+U yield h-site adsorption, where the $C_{6v}$ symmetry is broken by the NiH bond. For hydrogenated Co, we obtain h-site adsorption in GGA but GGA+U predicting t-site adsorption. For CoH, the adsorption heights of the transition metal atoms of $1.7\,\ang$ (at h-site, GGA) and $2.1\,\ang$ (t-site, GGA+U) are very close to the heights of Co in the respective high spin GGA+U structures. This gives a first indication that extended Co $4s$ derived states become partially occupied upon hydrogenation. Similarly, for NiH the metal atom heights of $1.7\,\ang$ ($1.8\,\ang$) above the h-site obtained from GGA (GGA+U) increase compared to the heights of isolated Ni atoms in low-spin configuration on graphene.
\begin{figure*}
\centering
\includegraphics[width=0.8\linewidth]{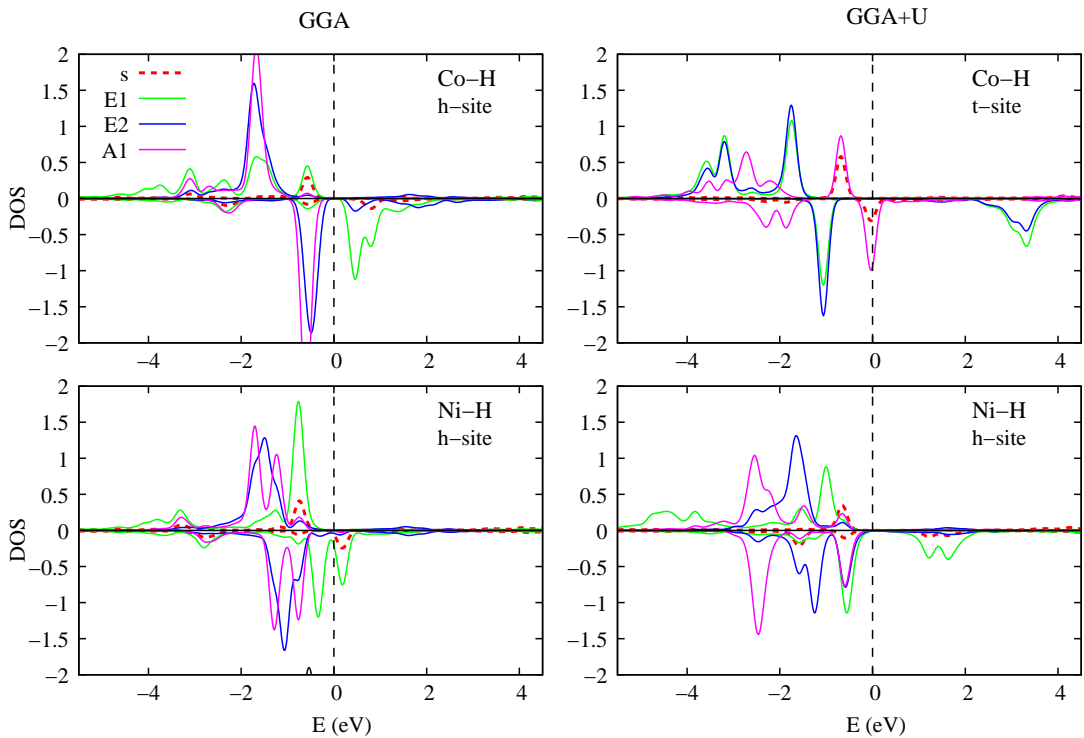}
\caption{(Color online) Local orbitally and spin resolved density of states for the hydrogenated adatoms CoH and NiH on graphene (from top to bottom) obtained with GGA (left column) as well as GGA+U (right column, $U=4$\,eV, $J=0.9$\,eV). Spin-up is plotted on the positive ordinate, spin-down on the negative ordinate.}
\label{fig:DOS_GGA_GGA+U_H-adatoms}
\end{figure*}
Independently of the employed functional (GGA or GGA+U), the supercell magnetic moments are $2.0\,\mu_B$ and $1.0\,\mu_B$ for CoH and NiH on graphene, respectively.

Well in line with the structural and magnetic changes upon hydrogenation, the LDOS of the Co $3d$ electrons in CoH is very similar to the high-spin solutions obtained for Co on graphene in GGA+U. The h-site configuration obtained for CoH in GGA has a $d$-electron LDOS which is very similar to the high-spin h-site solution for isolated Co on graphene (c.f. Ref. \onlinecite{CoGraphene_Kondo_PRB10}). The t-site GGA+U LDOS of CoH is similar to the LDOS of isolated Co at a t-site (Fig. \ref{fig:DOS_GGA_GGA+U_adatoms}). The major difference between hydrogenated and isolated Co on graphene is that there are no mainly $4s$-derived states pinned to the Fermi level in the former case.

For NiH, the LDOS (Fig. \ref{fig:DOS_GGA_GGA+U_H-adatoms}) shows that its magnetic moment is due to a hole in an orbital with large Ni $d_{xz},d_{yz}$-character which would correspond to orbitals transforming according to $E_1$ in a fully $C_{6v}$ symmetric environment. Here, $C_{6v}$ symmetry is broken by the NiH bond (Fig. \ref{fig:CoH_NiH_struct}). Analyzing the occupation matrix of the Ni $3d$-subspace as defined by the PAW basis sets we find 9 eigenvalues $>0.84$, i.e. 9 nearly filled Ni d-orbitals. One significantly lower eigenvalue of $0.42$ (GGA) as well as $0.22$ (GGA+U) corroborates that the hole leading to the supercell magnetic moment has largely Ni $3d$ character. The projection $\mathcal{P}_{E1}$ of the eigenvector $\ket{\Psi_0}$ belonging to this eigenvalue onto the Ni $d_{xz},d_{yz}$-subspace yields $\bra{\Psi_0}\mathcal{P}_{E1}\ket{\Psi_0}=0.96$ in GGA and $0.97$ in GGA+U. The hole in the Ni $3d$ derived states has consequently the major contribution from $E_1$ type orbitals. Hence, if NiH on graphene gives rise to a Kondo effect, the the orbitally controlled Kondo effect discussed for Co on graphene in Ref. \onlinecite{CoGraphene_Kondo_PRB10} likely generalizes the case at hand.

\section{Conclusions}
\label{sec:conclusions}
Based on first-principles calculations we discussed the interaction of graphene with transition metal adatoms. On-site Coulomb interactions can decisively affect interorbital charge transfers and thus control the electronic configuration, the adsorption geometry and the stability, particularly migration barriers, of adatoms on graphene.


Ad-atoms with high-spin ground state and (partially) occupied $4s$-orbital exhibit, in general, migration barriers on the order of $<0.06$\,eV. In contrast, adatoms with low-spin ground states and mainly empty $4s$-orbital are stabilized by covalent bonds of the $3d$-electrons to graphene and have significantly larger migration barriers of $0.15-0.5$\,eV. The thermal stability of adatoms can be probed, e.g., by scanning tunneling microscopy at different temperatures and will allow to draw conclusions on the electronic configuration of the adatoms. This is particularly interesting for the cases of Fe, Co, and Ni where GGA and GGA+U predict different ground state configurations.

The coexistence of high-spin and low spin solutions for Fe, Co, and Ni can be traced back to the fact that the $4s$-orbital of the adatom is usually chemically active but ``frustrated'' on graphene: For $3d$-transition metal adatoms on the normal metal surfaces the $4s$-electrons are fully delocalized and basically part of the conduction electron sea, whearas for adatoms on graphene a half-filled $4s$-orbital will be part of the impurity spin. 

Hydrogentation resolves the frustration of the $4s$-electrons and allows to tune magnetism of transition metal atoms on graphene chemically: NiH exhibits spin $S=1/2$ on graphene, while isolated Ni adatoms are nonmagnetic in GGA. Whether high- or low spin states are realized for a given transition metal adatom on graphene can be expected to depend very generally on the chemical environment. This issue including the influence of different substrates, e.g. insulators like SiO$_2$ \cite{Novoselov_science2004,Co_graphene_SiO2} or metals like Ir \cite{Michely_Ir_islands_graphene_PRL06,Ir_pinning_Feibelman_PRB08}, deserves future investigations.






\section*{Acknowledgments}
The authors are thankful to A. Balatsky, R. Berndt, K. Fauth, A. Geim, I. Grigorieva, M. Gyamfi, J. Honolka, J. Kr{\"o}ger, H. Manoharan, A. Rosch, M. Wasniowska, and  R. Wiesendanger for fruitful discussions. Support from SPP 1459, SFB 668 (Germany), the Cluster of Excellence ``Nanospintronics'' (LExI Hamburg), FOM (The Netherlands) as well as computer time from HLRN are acknowledged.

\bibliography{graphene_mag}

\begin{thebibliography}{34}%
\makeatletter
\providecommand \@ifxundefined [1]{%
 \@ifx{#1\undefined}
}%
\providecommand \@ifnum [1]{%
 \ifnum #1\expandafter \@firstoftwo
 \else \expandafter \@secondoftwo
 \fi
}%
\providecommand \@ifx [1]{%
 \ifx #1\expandafter \@firstoftwo
 \else \expandafter \@secondoftwo
 \fi
}%
\providecommand \natexlab [1]{#1}%
\providecommand \enquote  [1]{``#1''}%
\providecommand \bibnamefont  [1]{#1}%
\providecommand \bibfnamefont [1]{#1}%
\providecommand \citenamefont [1]{#1}%
\providecommand \href@noop [0]{\@secondoftwo}%
\providecommand \href [0]{\begingroup \@sanitize@url \@href}%
\providecommand \@href[1]{\@@startlink{#1}\@@href}%
\providecommand \@@href[1]{\endgroup#1\@@endlink}%
\providecommand \@sanitize@url [0]{\catcode `\\12\catcode `\$12\catcode
  `\&12\catcode `\#12\catcode `\^12\catcode `\_12\catcode `\%12\relax}%
\providecommand \@@startlink[1]{}%
\providecommand \@@endlink[0]{}%
\providecommand \url  [0]{\begingroup\@sanitize@url \@url }%
\providecommand \@url [1]{\endgroup\@href {#1}{\urlprefix }}%
\providecommand \urlprefix  [0]{URL }%
\providecommand \Eprint [0]{\href }%
\providecommand \doibase [0]{http://dx.doi.org/}%
\providecommand \selectlanguage [0]{\@gobble}%
\providecommand \bibinfo  [0]{\@secondoftwo}%
\providecommand \bibfield  [0]{\@secondoftwo}%
\providecommand \translation [1]{[#1]}%
\providecommand \BibitemOpen [0]{}%
\providecommand \bibitemStop [0]{}%
\providecommand \bibitemNoStop [0]{.\EOS\space}%
\providecommand \EOS [0]{\spacefactor3000\relax}%
\providecommand \BibitemShut  [1]{\csname bibitem#1\endcsname}%
\let\auto@bib@innerbib\@empty
\bibitem [{\citenamefont {Wallace}(1947)}]{Wallace-1947}%
  \BibitemOpen
  \bibfield  {author} {\bibinfo {author} {\bibfnamefont {P.~R.}\ \bibnamefont
  {Wallace}},\ }\href@noop {} {\bibfield  {journal} {\bibinfo  {journal} {Phys.
  Rev.}\ }\textbf {\bibinfo {volume} {71}},\ \bibinfo {pages} {622} (\bibinfo
  {year} {1947})}\BibitemShut {NoStop}%
\bibitem [{\citenamefont {McClure}(1956)}]{McClure_56}%
  \BibitemOpen
  \bibfield  {author} {\bibinfo {author} {\bibfnamefont {J.~W.}\ \bibnamefont
  {McClure}},\ }\href@noop {} {\bibfield  {journal} {\bibinfo  {journal} {Phys.
  Rev.}\ }\textbf {\bibinfo {volume} {104}},\ \bibinfo {pages} {666} (\bibinfo
  {year} {1956})}\BibitemShut {NoStop}%
\bibitem [{\citenamefont {Novoselov}\ \emph {et~al.}(2004)\citenamefont
  {Novoselov}, \citenamefont {Geim}, \citenamefont {Morozov}, \citenamefont
  {Jiang}, \citenamefont {Zhang}, \citenamefont {Dubonos}, \citenamefont
  {Grigorieva},\ and\ \citenamefont {Firsov}}]{Novoselov_science2004}%
  \BibitemOpen
  \bibfield  {author} {\bibinfo {author} {\bibfnamefont {K.~S.}\ \bibnamefont
  {Novoselov}}, \bibinfo {author} {\bibfnamefont {A.~K.}\ \bibnamefont {Geim}},
  \bibinfo {author} {\bibfnamefont {S.~V.}\ \bibnamefont {Morozov}}, \bibinfo
  {author} {\bibfnamefont {D.}~\bibnamefont {Jiang}}, \bibinfo {author}
  {\bibfnamefont {Y.}~\bibnamefont {Zhang}}, \bibinfo {author} {\bibfnamefont
  {S.~V.}\ \bibnamefont {Dubonos}}, \bibinfo {author} {\bibfnamefont {I.~V.}\
  \bibnamefont {Grigorieva}}, \ and\ \bibinfo {author} {\bibfnamefont {A.~A.}\
  \bibnamefont {Firsov}},\ }\href@noop {} {\bibfield  {journal} {\bibinfo
  {journal} {Science}\ }\textbf {\bibinfo {volume} {306}},\ \bibinfo {pages}
  {666} (\bibinfo {year} {2004})}\BibitemShut {NoStop}%
\bibitem [{\citenamefont {Schedin}\ \emph {et~al.}(2007)\citenamefont
  {Schedin}, \citenamefont {Geim}, \citenamefont {Morozov}, \citenamefont
  {Hill}, \citenamefont {Blake}, \citenamefont {Katsnelson},\ and\
  \citenamefont {Novoselov}}]{SchedinGassensors}%
  \BibitemOpen
  \bibfield  {author} {\bibinfo {author} {\bibfnamefont {F.}~\bibnamefont
  {Schedin}}, \bibinfo {author} {\bibfnamefont {A.~K.}\ \bibnamefont {Geim}},
  \bibinfo {author} {\bibfnamefont {S.~V.}\ \bibnamefont {Morozov}}, \bibinfo
  {author} {\bibfnamefont {E.~W.}\ \bibnamefont {Hill}}, \bibinfo {author}
  {\bibfnamefont {P.}~\bibnamefont {Blake}}, \bibinfo {author} {\bibfnamefont
  {M.~I.}\ \bibnamefont {Katsnelson}}, \ and\ \bibinfo {author} {\bibfnamefont
  {K.~S.}\ \bibnamefont {Novoselov}},\ }\href@noop {} {\bibfield  {journal}
  {\bibinfo  {journal} {Nat. Mater.}\ }\textbf {\bibinfo {volume} {6}},\
  \bibinfo {pages} {652} (\bibinfo {year} {2007})}\BibitemShut {NoStop}%
\bibitem [{\citenamefont {Wehling}\ \emph {et~al.}(2008)\citenamefont
  {Wehling}, \citenamefont {Novoselov}, \citenamefont {Morozov}, \citenamefont
  {Vdovin}, \citenamefont {Katsnelson}, \citenamefont {Geim},\ and\
  \citenamefont {Lichtenstein}}]{NanoLettAds}%
  \BibitemOpen
  \bibfield  {author} {\bibinfo {author} {\bibfnamefont {T.~O.}\ \bibnamefont
  {Wehling}}, \bibinfo {author} {\bibfnamefont {K.~S.}\ \bibnamefont
  {Novoselov}}, \bibinfo {author} {\bibfnamefont {S.~V.}\ \bibnamefont
  {Morozov}}, \bibinfo {author} {\bibfnamefont {E.~E.}\ \bibnamefont {Vdovin}},
  \bibinfo {author} {\bibfnamefont {M.~I.}\ \bibnamefont {Katsnelson}},
  \bibinfo {author} {\bibfnamefont {A.~K.}\ \bibnamefont {Geim}}, \ and\
  \bibinfo {author} {\bibfnamefont {A.~I.}\ \bibnamefont {Lichtenstein}},\
  }\href@noop {} {\bibfield  {journal} {\bibinfo  {journal} {Nano Lett.}\
  }\textbf {\bibinfo {volume} {8}},\ \bibinfo {pages} {173} (\bibinfo {year}
  {2008})}\BibitemShut {NoStop}%
\bibitem [{\citenamefont {Chen}\ \emph {et~al.}(2008)\citenamefont {Chen},
  \citenamefont {Jang}, \citenamefont {Adam}, \citenamefont {Fuhrer},
  \citenamefont {Williams},\ and\ \citenamefont
  {Ishigami}}]{AdamFuhrerIshigamiCargedImp08}%
  \BibitemOpen
  \bibfield  {author} {\bibinfo {author} {\bibfnamefont {J.~H.}\ \bibnamefont
  {Chen}}, \bibinfo {author} {\bibfnamefont {C.}~\bibnamefont {Jang}}, \bibinfo
  {author} {\bibfnamefont {S.}~\bibnamefont {Adam}}, \bibinfo {author}
  {\bibfnamefont {M.~S.}\ \bibnamefont {Fuhrer}}, \bibinfo {author}
  {\bibfnamefont {E.~D.}\ \bibnamefont {Williams}}, \ and\ \bibinfo {author}
  {\bibfnamefont {M.}~\bibnamefont {Ishigami}},\ }\href@noop {} {\bibfield
  {journal} {\bibinfo  {journal} {Nature Phys.}\ }\textbf {\bibinfo {volume}
  {4}},\ \bibinfo {pages} {377} (\bibinfo {year} {2008})}\BibitemShut {NoStop}%
\bibitem [{\citenamefont {Yagi}\ \emph {et~al.}(2004)\citenamefont {Yagi},
  \citenamefont {Briere}, \citenamefont {Sluiter}, \citenamefont {Kumar},
  \citenamefont {Farajian},\ and\ \citenamefont {Kawazoe}}]{Kawazoe04}%
  \BibitemOpen
  \bibfield  {author} {\bibinfo {author} {\bibfnamefont {Y.}~\bibnamefont
  {Yagi}}, \bibinfo {author} {\bibfnamefont {T.~M.}\ \bibnamefont {Briere}},
  \bibinfo {author} {\bibfnamefont {M.~H.~F.}\ \bibnamefont {Sluiter}},
  \bibinfo {author} {\bibfnamefont {V.}~\bibnamefont {Kumar}}, \bibinfo
  {author} {\bibfnamefont {A.~A.}\ \bibnamefont {Farajian}}, \ and\ \bibinfo
  {author} {\bibfnamefont {Y.}~\bibnamefont {Kawazoe}},\ }\href {\doibase
  10.1103/PhysRevB.69.075414} {\bibfield  {journal} {\bibinfo  {journal} {Phys.
  Rev. B}\ }\textbf {\bibinfo {volume} {69}},\ \bibinfo {pages} {075414}
  (\bibinfo {year} {2004})}\BibitemShut {NoStop}%
\bibitem [{\citenamefont {Chan}\ \emph {et~al.}(2008)\citenamefont {Chan},
  \citenamefont {Neaton},\ and\ \citenamefont {Cohen}}]{Cohen_PRB08}%
  \BibitemOpen
  \bibfield  {author} {\bibinfo {author} {\bibfnamefont {K.~T.}\ \bibnamefont
  {Chan}}, \bibinfo {author} {\bibfnamefont {J.~B.}\ \bibnamefont {Neaton}}, \
  and\ \bibinfo {author} {\bibfnamefont {M.~L.}\ \bibnamefont {Cohen}},\ }\href
  {\doibase 10.1103/PhysRevB.77.235430} {\bibfield  {journal} {\bibinfo
  {journal} {Phys. Rev. B}\ }\textbf {\bibinfo {volume} {77}},\ \bibinfo
  {pages} {235430} (\bibinfo {year} {2008})}\BibitemShut {NoStop}%
\bibitem [{\citenamefont {Mao}\ \emph {et~al.}(2008)\citenamefont {Mao},
  \citenamefont {Yuan},\ and\ \citenamefont {Zhong}}]{Mao_Co_graphene_09}%
  \BibitemOpen
  \bibfield  {author} {\bibinfo {author} {\bibfnamefont {Y.}~\bibnamefont
  {Mao}}, \bibinfo {author} {\bibfnamefont {J.}~\bibnamefont {Yuan}}, \ and\
  \bibinfo {author} {\bibfnamefont {J.}~\bibnamefont {Zhong}},\ }\href@noop {}
  {\bibfield  {journal} {\bibinfo  {journal} {J. Phys.: Condens. Matter}\
  }\textbf {\bibinfo {volume} {20}},\ \bibinfo {pages} {115209} (\bibinfo
  {year} {2008})}\BibitemShut {NoStop}%
\bibitem [{\citenamefont {Johll}\ \emph {et~al.}(2009)\citenamefont {Johll},
  \citenamefont {Kang},\ and\ \citenamefont {Tok}}]{Tok_PRB09}%
  \BibitemOpen
  \bibfield  {author} {\bibinfo {author} {\bibfnamefont {H.}~\bibnamefont
  {Johll}}, \bibinfo {author} {\bibfnamefont {H.~C.}\ \bibnamefont {Kang}}, \
  and\ \bibinfo {author} {\bibfnamefont {E.~S.}\ \bibnamefont {Tok}},\ }\href
  {\doibase 10.1103/PhysRevB.79.245416} {\bibfield  {journal} {\bibinfo
  {journal} {Phys. Rev. B}\ }\textbf {\bibinfo {volume} {79}},\ \bibinfo
  {pages} {245416} (\bibinfo {year} {2009})}\BibitemShut {NoStop}%
\bibitem [{\citenamefont {Longo}\ \emph {et~al.}(2010)\citenamefont {Longo},
  \citenamefont {Carrete}, \citenamefont {Ferrer},\ and\ \citenamefont
  {Gallego}}]{Longo_PRB10}%
  \BibitemOpen
  \bibfield  {author} {\bibinfo {author} {\bibfnamefont {R.~C.}\ \bibnamefont
  {Longo}}, \bibinfo {author} {\bibfnamefont {J.}~\bibnamefont {Carrete}},
  \bibinfo {author} {\bibfnamefont {J.}~\bibnamefont {Ferrer}}, \ and\ \bibinfo
  {author} {\bibfnamefont {L.~J.}\ \bibnamefont {Gallego}},\ }\href {\doibase
  10.1103/PhysRevB.81.115418} {\bibfield  {journal} {\bibinfo  {journal} {Phys.
  Rev. B}\ }\textbf {\bibinfo {volume} {81}},\ \bibinfo {pages} {115418}
  (\bibinfo {year} {2010})}\BibitemShut {NoStop}%
\bibitem [{\citenamefont {Wehling}\ \emph
  {et~al.}(2010{\natexlab{a}})\citenamefont {Wehling}, \citenamefont {Dahal},
  \citenamefont {Lichtenstein}, \citenamefont {Katsnelson}, \citenamefont
  {Manoharan},\ and\ \citenamefont {Balatsky}}]{Co_Fano_PRB10}%
  \BibitemOpen
  \bibfield  {author} {\bibinfo {author} {\bibfnamefont {T.~O.}\ \bibnamefont
  {Wehling}}, \bibinfo {author} {\bibfnamefont {H.~P.}\ \bibnamefont {Dahal}},
  \bibinfo {author} {\bibfnamefont {A.~I.}\ \bibnamefont {Lichtenstein}},
  \bibinfo {author} {\bibfnamefont {M.~I.}\ \bibnamefont {Katsnelson}},
  \bibinfo {author} {\bibfnamefont {H.~C.}\ \bibnamefont {Manoharan}}, \ and\
  \bibinfo {author} {\bibfnamefont {A.~V.}\ \bibnamefont {Balatsky}},\ }\href
  {\doibase 10.1103/PhysRevB.81.085413} {\bibfield  {journal} {\bibinfo
  {journal} {Phys. Rev. B}\ }\textbf {\bibinfo {volume} {81}},\ \bibinfo
  {pages} {085413} (\bibinfo {year} {2010}{\natexlab{a}})}\BibitemShut
  {NoStop}%
\bibitem [{\citenamefont {Cao}\ \emph {et~al.}(2010)\citenamefont {Cao},
  \citenamefont {Wu}, \citenamefont {Jiang},\ and\ \citenamefont
  {Cheng}}]{Cao_PRB10}%
  \BibitemOpen
  \bibfield  {author} {\bibinfo {author} {\bibfnamefont {C.}~\bibnamefont
  {Cao}}, \bibinfo {author} {\bibfnamefont {M.}~\bibnamefont {Wu}}, \bibinfo
  {author} {\bibfnamefont {J.}~\bibnamefont {Jiang}}, \ and\ \bibinfo {author}
  {\bibfnamefont {H.-P.}\ \bibnamefont {Cheng}},\ }\href {\doibase
  10.1103/PhysRevB.81.205424} {\bibfield  {journal} {\bibinfo  {journal} {Phys.
  Rev. B}\ }\textbf {\bibinfo {volume} {81}},\ \bibinfo {pages} {205424}
  (\bibinfo {year} {2010})}\BibitemShut {NoStop}%
\bibitem [{\citenamefont {Wehling}\ \emph
  {et~al.}(2010{\natexlab{b}})\citenamefont {Wehling}, \citenamefont
  {Balatsky}, \citenamefont {Katsnelson}, \citenamefont {Lichtenstein},\ and\
  \citenamefont {Rosch}}]{CoGraphene_Kondo_PRB10}%
  \BibitemOpen
  \bibfield  {author} {\bibinfo {author} {\bibfnamefont {T.~O.}\ \bibnamefont
  {Wehling}}, \bibinfo {author} {\bibfnamefont {A.~V.}\ \bibnamefont
  {Balatsky}}, \bibinfo {author} {\bibfnamefont {M.~I.}\ \bibnamefont
  {Katsnelson}}, \bibinfo {author} {\bibfnamefont {A.~I.}\ \bibnamefont
  {Lichtenstein}}, \ and\ \bibinfo {author} {\bibfnamefont {A.}~\bibnamefont
  {Rosch}},\ }\href {\doibase 10.1103/PhysRevB.81.115427} {\bibfield  {journal}
  {\bibinfo  {journal} {Phys. Rev. B}\ }\textbf {\bibinfo {volume} {81}},\
  \bibinfo {pages} {115427} (\bibinfo {year} {2010}{\natexlab{b}})}\BibitemShut
  {NoStop}%
\bibitem [{\citenamefont {Yazyev}\ and\ \citenamefont
  {Pasquarello}(2010)}]{Yazyev_PRB10}%
  \BibitemOpen
  \bibfield  {author} {\bibinfo {author} {\bibfnamefont {O.~V.}\ \bibnamefont
  {Yazyev}}\ and\ \bibinfo {author} {\bibfnamefont {A.}~\bibnamefont
  {Pasquarello}},\ }\href {\doibase 10.1103/PhysRevB.82.045407} {\bibfield
  {journal} {\bibinfo  {journal} {Phys. Rev. B}\ }\textbf {\bibinfo {volume}
  {82}},\ \bibinfo {pages} {045407} (\bibinfo {year} {2010})}\BibitemShut
  {NoStop}%
\bibitem [{\citenamefont {Jacob}\ and\ \citenamefont
  {Kotliar}(2010)}]{Jacob_Co_Graphene10}%
  \BibitemOpen
  \bibfield  {author} {\bibinfo {author} {\bibfnamefont {D.}~\bibnamefont
  {Jacob}}\ and\ \bibinfo {author} {\bibfnamefont {G.}~\bibnamefont
  {Kotliar}},\ }\href {\doibase 10.1103/PhysRevB.82.085423} {\bibfield
  {journal} {\bibinfo  {journal} {Phys. Rev. B}\ }\textbf {\bibinfo {volume}
  {82}},\ \bibinfo {pages} {085423} (\bibinfo {year} {2010})}\BibitemShut
  {NoStop}%
\bibitem [{\citenamefont {Chan}\ \emph {et~al.}(2011)\citenamefont {Chan},
  \citenamefont {Lee},\ and\ \citenamefont {Cohen}}]{Co_adat_CohenPRB11}%
  \BibitemOpen
  \bibfield  {author} {\bibinfo {author} {\bibfnamefont {K.~T.}\ \bibnamefont
  {Chan}}, \bibinfo {author} {\bibfnamefont {H.}~\bibnamefont {Lee}}, \ and\
  \bibinfo {author} {\bibfnamefont {M.~L.}\ \bibnamefont {Cohen}},\ }\href
  {\doibase 10.1103/PhysRevB.83.035405} {\bibfield  {journal} {\bibinfo
  {journal} {Phys. Rev. B}\ }\textbf {\bibinfo {volume} {83}},\ \bibinfo
  {pages} {035405} (\bibinfo {year} {2011})}\BibitemShut {NoStop}%
\bibitem [{\citenamefont {Liu}\ \emph {et~al.}(2011)\citenamefont {Liu},
  \citenamefont {Wang}, \citenamefont {Yao}, \citenamefont {Lu}, \citenamefont
  {Hupalo}, \citenamefont {Tringides},\ and\ \citenamefont {Ho}}]{Ho_PRB11}%
  \BibitemOpen
  \bibfield  {author} {\bibinfo {author} {\bibfnamefont {X.}~\bibnamefont
  {Liu}}, \bibinfo {author} {\bibfnamefont {C.~Z.}\ \bibnamefont {Wang}},
  \bibinfo {author} {\bibfnamefont {Y.~X.}\ \bibnamefont {Yao}}, \bibinfo
  {author} {\bibfnamefont {W.~C.}\ \bibnamefont {Lu}}, \bibinfo {author}
  {\bibfnamefont {M.}~\bibnamefont {Hupalo}}, \bibinfo {author} {\bibfnamefont
  {M.~C.}\ \bibnamefont {Tringides}}, \ and\ \bibinfo {author} {\bibfnamefont
  {K.~M.}\ \bibnamefont {Ho}},\ }\href {\doibase 10.1103/PhysRevB.83.235411}
  {\bibfield  {journal} {\bibinfo  {journal} {Phys. Rev. B}\ }\textbf {\bibinfo
  {volume} {83}},\ \bibinfo {pages} {235411} (\bibinfo {year}
  {2011})}\BibitemShut {NoStop}%
\bibitem [{\citenamefont {Nakada}\ and\ \citenamefont
  {Ishii}(2011)}]{Nakada11}%
  \BibitemOpen
  \bibfield  {author} {\bibinfo {author} {\bibfnamefont {K.}~\bibnamefont
  {Nakada}}\ and\ \bibinfo {author} {\bibfnamefont {A.}~\bibnamefont {Ishii}},\
  }\href@noop {} {\bibfield  {journal} {\bibinfo  {journal} {Solid State Com.}\
  }\textbf {\bibinfo {volume} {151}},\ \bibinfo {pages} {13} (\bibinfo {year}
  {2011})}\BibitemShut {NoStop}%
\bibitem [{\citenamefont {Wehling}\ \emph {et~al.}(2009)\citenamefont
  {Wehling}, \citenamefont {Katsnelson},\ and\ \citenamefont
  {Lichtenstein}}]{MonoImp09}%
  \BibitemOpen
  \bibfield  {author} {\bibinfo {author} {\bibfnamefont {T.~O.}\ \bibnamefont
  {Wehling}}, \bibinfo {author} {\bibfnamefont {M.~I.}\ \bibnamefont
  {Katsnelson}}, \ and\ \bibinfo {author} {\bibfnamefont {A.~I.}\ \bibnamefont
  {Lichtenstein}},\ }\href@noop {} {\bibfield  {journal} {\bibinfo  {journal}
  {Phys. Rev. B}\ }\textbf {\bibinfo {volume} {80}},\ \bibinfo {pages} {085428}
  (\bibinfo {year} {2009})}\BibitemShut {NoStop}%
\bibitem [{\citenamefont {Kresse}\ and\ \citenamefont
  {Hafner}(1994)}]{Kresse:PP_VASP}%
  \BibitemOpen
  \bibfield  {author} {\bibinfo {author} {\bibfnamefont {G.}~\bibnamefont
  {Kresse}}\ and\ \bibinfo {author} {\bibfnamefont {J.}~\bibnamefont
  {Hafner}},\ }\href@noop {} {\bibfield  {journal} {\bibinfo  {journal} {J.
  Phys.: Condes. Matter}\ }\textbf {\bibinfo {volume} {6}},\ \bibinfo {pages}
  {8245} (\bibinfo {year} {1994})}\BibitemShut {NoStop}%
\bibitem [{\citenamefont {Bl\"ochl}(1994)}]{Bloechl:PAW1994}%
  \BibitemOpen
  \bibfield  {author} {\bibinfo {author} {\bibfnamefont {P.~E.}\ \bibnamefont
  {Bl\"ochl}},\ }\href@noop {} {\bibfield  {journal} {\bibinfo  {journal}
  {Phys. Rev. B}\ }\textbf {\bibinfo {volume} {50}},\ \bibinfo {pages} {17953}
  (\bibinfo {year} {1994})}\BibitemShut {NoStop}%
\bibitem [{\citenamefont {Kresse}\ and\ \citenamefont
  {Joubert}(1999)}]{Kresse:PAW_VASP}%
  \BibitemOpen
  \bibfield  {author} {\bibinfo {author} {\bibfnamefont {G.}~\bibnamefont
  {Kresse}}\ and\ \bibinfo {author} {\bibfnamefont {D.}~\bibnamefont
  {Joubert}},\ }\href@noop {} {\bibfield  {journal} {\bibinfo  {journal} {Phys.
  Rev. B}\ }\textbf {\bibinfo {volume} {59}},\ \bibinfo {pages} {1758}
  (\bibinfo {year} {1999})}\BibitemShut {NoStop}%
\bibitem [{\citenamefont {Perdew}\ \emph {et~al.}(1992)\citenamefont {Perdew},
  \citenamefont {Chevary}, \citenamefont {Vosko}, \citenamefont {Jackson},
  \citenamefont {Pederson}, \citenamefont {Singh},\ and\ \citenamefont
  {Fiolhais}}]{Perdew:PW91}%
  \BibitemOpen
  \bibfield  {author} {\bibinfo {author} {\bibfnamefont {J.~P.}\ \bibnamefont
  {Perdew}}, \bibinfo {author} {\bibfnamefont {J.~A.}\ \bibnamefont {Chevary}},
  \bibinfo {author} {\bibfnamefont {S.~H.}\ \bibnamefont {Vosko}}, \bibinfo
  {author} {\bibfnamefont {K.~A.}\ \bibnamefont {Jackson}}, \bibinfo {author}
  {\bibfnamefont {M.~R.}\ \bibnamefont {Pederson}}, \bibinfo {author}
  {\bibfnamefont {D.~J.}\ \bibnamefont {Singh}}, \ and\ \bibinfo {author}
  {\bibfnamefont {C.}~\bibnamefont {Fiolhais}},\ }\href@noop {} {\bibfield
  {journal} {\bibinfo  {journal} {Phys. Rev. B}\ }\textbf {\bibinfo {volume}
  {46}},\ \bibinfo {pages} {6671} (\bibinfo {year} {1992})}\BibitemShut
  {NoStop}%
\bibitem [{\citenamefont {Solovyev}\ \emph {et~al.}(1994)\citenamefont
  {Solovyev}, \citenamefont {Dederichs},\ and\ \citenamefont
  {Anisimov}}]{Anisimov_PRB94}%
  \BibitemOpen
  \bibfield  {author} {\bibinfo {author} {\bibfnamefont {I.~V.}\ \bibnamefont
  {Solovyev}}, \bibinfo {author} {\bibfnamefont {P.~H.}\ \bibnamefont
  {Dederichs}}, \ and\ \bibinfo {author} {\bibfnamefont {V.~I.}\ \bibnamefont
  {Anisimov}},\ }\href@noop {} {\bibfield  {journal} {\bibinfo  {journal}
  {Phys. Rev. B}\ }\textbf {\bibinfo {volume} {50}},\ \bibinfo {pages} {16861}
  (\bibinfo {year} {1994})}\BibitemShut {NoStop}%
\bibitem [{\citenamefont {Anisimov}\ \emph {et~al.}(1997)\citenamefont
  {Anisimov}, \citenamefont {Aryasetiawan},\ and\ \citenamefont
  {Lichtenstein}}]{Anisimov_Lichtenstein_LDAU_97}%
  \BibitemOpen
  \bibfield  {author} {\bibinfo {author} {\bibfnamefont {V.~I.}\ \bibnamefont
  {Anisimov}}, \bibinfo {author} {\bibfnamefont {F.}~\bibnamefont
  {Aryasetiawan}}, \ and\ \bibinfo {author} {\bibfnamefont {A.~I.}\
  \bibnamefont {Lichtenstein}},\ }\href@noop {} {\bibfield  {journal} {\bibinfo
   {journal} {J. Phys.: Condens. Matter}\ }\textbf {\bibinfo {volume} {9}},\
  \bibinfo {pages} {767} (\bibinfo {year} {1997})}\BibitemShut {NoStop}%
\bibitem [{Note1()}]{Note1}%
  \BibitemOpen
  \bibinfo {note} {The paths were obtained by linearly interpolating between
  the relaxed adsorption geometries and using 5 images in between each of the
  relaxed geometries.}\BibitemShut {Stop}%
\bibitem [{\citenamefont {Moroni}\ \emph {et~al.}(1997)\citenamefont {Moroni},
  \citenamefont {Kresse}, \citenamefont {Hafner},\ and\ \citenamefont
  {Furthm\"uller}}]{Kresse_PRB97}%
  \BibitemOpen
  \bibfield  {author} {\bibinfo {author} {\bibfnamefont {E.~G.}\ \bibnamefont
  {Moroni}}, \bibinfo {author} {\bibfnamefont {G.}~\bibnamefont {Kresse}},
  \bibinfo {author} {\bibfnamefont {J.}~\bibnamefont {Hafner}}, \ and\ \bibinfo
  {author} {\bibfnamefont {J.}~\bibnamefont {Furthm\"uller}},\ }\href {\doibase
  10.1103/PhysRevB.56.15629} {\bibfield  {journal} {\bibinfo  {journal} {Phys.
  Rev. B}\ }\textbf {\bibinfo {volume} {56}},\ \bibinfo {pages} {15629}
  (\bibinfo {year} {1997})}\BibitemShut {NoStop}%
\bibitem [{Note2()}]{Note2}%
  \BibitemOpen
  \bibinfo {note} {The exact amount of charge transfer can depend on the adatom
  concentration / supercell size. The numbers given, here, should be considered
  qualitatively.}\BibitemShut {Stop}%
\bibitem [{\citenamefont {N\'{e}el}\ \emph {et~al.}(2008)\citenamefont
  {N\'{e}el}, \citenamefont {Kr\"{o}ger}, \citenamefont {Berndt}, \citenamefont
  {Wehling}, \citenamefont {Lichtenstein},\ and\ \citenamefont
  {Katsnelson}}]{CoCu_n_PRL08}%
  \BibitemOpen
  \bibfield  {author} {\bibinfo {author} {\bibfnamefont {N.}~\bibnamefont
  {N\'{e}el}}, \bibinfo {author} {\bibfnamefont {J.}~\bibnamefont
  {Kr\"{o}ger}}, \bibinfo {author} {\bibfnamefont {R.}~\bibnamefont {Berndt}},
  \bibinfo {author} {\bibfnamefont {T.~O.}\ \bibnamefont {Wehling}}, \bibinfo
  {author} {\bibfnamefont {A.~I.}\ \bibnamefont {Lichtenstein}}, \ and\
  \bibinfo {author} {\bibfnamefont {M.~I.}\ \bibnamefont {Katsnelson}},\
  }\href@noop {} {\bibfield  {journal} {\bibinfo  {journal} {Phys. Rev. Lett.}\
  }\textbf {\bibinfo {volume} {101}},\ \bibinfo {pages} {266803} (\bibinfo
  {year} {2008})}\BibitemShut {NoStop}%
\bibitem [{\citenamefont {Rudenko}\ \emph {et~al.}(2009)\citenamefont
  {Rudenko}, \citenamefont {Mazurenko}, \citenamefont {Anisimov},\ and\
  \citenamefont {Lichtenstein}}]{Rudenko_09}%
  \BibitemOpen
  \bibfield  {author} {\bibinfo {author} {\bibfnamefont {A.~N.}\ \bibnamefont
  {Rudenko}}, \bibinfo {author} {\bibfnamefont {V.~V.}\ \bibnamefont
  {Mazurenko}}, \bibinfo {author} {\bibfnamefont {V.~I.}\ \bibnamefont
  {Anisimov}}, \ and\ \bibinfo {author} {\bibfnamefont {A.~I.}\ \bibnamefont
  {Lichtenstein}},\ }\href@noop {} {\bibfield  {journal} {\bibinfo  {journal}
  {Phys. Rev. B}\ }\textbf {\bibinfo {volume} {79}},\ \bibinfo {pages} {144418}
  (\bibinfo {year} {2009})}\BibitemShut {NoStop}%
\bibitem [{\citenamefont {Brar}\ \emph {et~al.}(2011)\citenamefont {Brar},
  \citenamefont {Decker}, \citenamefont {Solowan}, \citenamefont {Wang},
  \citenamefont {Maserati}, \citenamefont {Chan}, \citenamefont {Lee},
  \citenamefont {Girit}, \citenamefont {Zettl}, \citenamefont {Louie},
  \citenamefont {Cohen},\ and\ \citenamefont {Crommie}}]{Co_graphene_SiO2}%
  \BibitemOpen
  \bibfield  {author} {\bibinfo {author} {\bibfnamefont {V.~W.}\ \bibnamefont
  {Brar}}, \bibinfo {author} {\bibfnamefont {R.}~\bibnamefont {Decker}},
  \bibinfo {author} {\bibfnamefont {H.-M.}\ \bibnamefont {Solowan}}, \bibinfo
  {author} {\bibfnamefont {Y.}~\bibnamefont {Wang}}, \bibinfo {author}
  {\bibfnamefont {L.}~\bibnamefont {Maserati}}, \bibinfo {author}
  {\bibfnamefont {K.~T.}\ \bibnamefont {Chan}}, \bibinfo {author}
  {\bibfnamefont {H.}~\bibnamefont {Lee}}, \bibinfo {author} {\bibfnamefont
  {C.~O.}\ \bibnamefont {Girit}}, \bibinfo {author} {\bibfnamefont
  {A.}~\bibnamefont {Zettl}}, \bibinfo {author} {\bibfnamefont {S.~G.}\
  \bibnamefont {Louie}}, \bibinfo {author} {\bibfnamefont {M.~L.}\ \bibnamefont
  {Cohen}}, \ and\ \bibinfo {author} {\bibfnamefont {M.~F.}\ \bibnamefont
  {Crommie}},\ }\href@noop {} {\bibfield  {journal} {\bibinfo  {journal}
  {Nature Phys.}\ }\textbf {\bibinfo {volume} {7}},\ \bibinfo {pages} {43}
  (\bibinfo {year} {2011})}\BibitemShut {NoStop}%
\bibitem [{\citenamefont {N'Diaye}\ \emph {et~al.}(2006)\citenamefont
  {N'Diaye}, \citenamefont {Bleikamp}, \citenamefont {Feibelman},\ and\
  \citenamefont {Michely}}]{Michely_Ir_islands_graphene_PRL06}%
  \BibitemOpen
  \bibfield  {author} {\bibinfo {author} {\bibfnamefont {A.~T.}\ \bibnamefont
  {N'Diaye}}, \bibinfo {author} {\bibfnamefont {S.}~\bibnamefont {Bleikamp}},
  \bibinfo {author} {\bibfnamefont {P.~J.}\ \bibnamefont {Feibelman}}, \ and\
  \bibinfo {author} {\bibfnamefont {T.}~\bibnamefont {Michely}},\ }\href
  {\doibase 10.1103/PhysRevLett.97.215501} {\bibfield  {journal} {\bibinfo
  {journal} {Phys. Rev. Lett.}\ }\textbf {\bibinfo {volume} {97}},\ \bibinfo
  {pages} {215501} (\bibinfo {year} {2006})}\BibitemShut {NoStop}%
\bibitem [{\citenamefont {Feibelman}(2008)}]{Ir_pinning_Feibelman_PRB08}%
  \BibitemOpen
  \bibfield  {author} {\bibinfo {author} {\bibfnamefont {P.~J.}\ \bibnamefont
  {Feibelman}},\ }\href {\doibase 10.1103/PhysRevB.77.165419} {\bibfield
  {journal} {\bibinfo  {journal} {Phys. Rev. B}\ }\textbf {\bibinfo {volume}
  {77}},\ \bibinfo {pages} {165419} (\bibinfo {year} {2008})}\BibitemShut
  {NoStop}%
\end{thebibliography}%

\end{document}